\documentclass[aps,twocolumn,prc,showpacs,superscriptaddress,floatfix]{revtex4}
\usepackage{graphicx}
\usepackage{epsf}
\usepackage{wrapfig}
\usepackage{epsfig}
\usepackage{sublabel}
\usepackage{epsfig}
\usepackage{verbatim}
\usepackage{amsmath}
\usepackage{color}
\usepackage{soul}
\def\Vec#1{\mbox{\boldmath $#1$}}

\def\beqy{\begin{eqnarray}}
\def\eeqy{\end{eqnarray}}

\newcommand\beq{\begin{equation}}

\newcommand\eeq{\end{equation}}
\newcommand\beqn{\begin{eqnarray}}
\newcommand\eeqn{\end{eqnarray}}

\newcommand{\la}{\langle}
\newcommand{\ra}{\rangle}

\def\Re{\,\mbox{Re}\,}

\def\sqelha{\sigma_{qel}^{hA}}
\def\sthn{\sigma_{tot}^{hN}}
\def\stha{\sigma_{tot}^{hA}}

\def\selha{\sigma_{el}^{hA}}
\def\sinhn{\sigma_{in}^{hN}}

\def\stap{\sigma_{tot}^{pA}}
\def\stota{\sigma_{tot}^{hA}}
\def\sela{\sigma_{el}^{hA}}
\def\sina{\sigma_{in}^{hA}}

\def\sinn{\sigma_{in}^{NN}}
\def\sinhn{\sigma_{in}^{hN}}
\def\sinha{\sigma_{in}^{hA}}

\begin{document}
\date{\today}
\vskip 2mm
\title{NUMBER OF COLLISIONS IN THE GLAUBER MODEL AND BEYOND}
\author{C. Ciofi degli Atti}
\affiliation{
   Istituto Nazionale di Fisica Nucleare, Sezione di Perugia,
c/o Department of Physics, University of Perugia,
   Via A. Pascoli, I-06123, Italy}
   \author{B. Z. Kopeliovich}
   \affiliation{Departamento de F\'{\i}sica,
Universidad T\'ecnica Federico Santa Mar\'{\i}a and
\\
Instituto de Estudios Avanzados en Ciencias e Ingenier\'{\i} \\
Centro Cient\'ifico-Tecnol\'ogico de Valpara\'iso;\\
Casilla 110-V, Valpara\'iso, Chile}
\author{C. B. Mezzetti}
\affiliation{Department of Physics, University of Perugia and
Istituto Nazionale di Fisica Nucleare, Sezione di Perugia,
   Via A. Pascoli, I-06123, Italy}
\author{I. K. Potashnikova}
\affiliation{Departamento de F\'{\i}sica,
Universidad T\'ecnica Federico Santa Mar\'{\i}a and
\\
Instituto de Estudios Avanzados en Ciencias e Ingenier\'{\i} \\
Centro Cient\'ifico-Tecnol\'ogico de Valpara\'iso;\\
Casilla 110-V, Valpara\'iso, Chile}
\author{I. Schmidt}
\affiliation{Departamento de F\'{\i}sica,
Universidad T\'ecnica Federico Santa Mar\'{\i}a and
\\
Instituto de Estudios Avanzados en Ciencias e Ingenier\'{\i} \\
Centro Cient\'ifico-Tecnol\'ogico de Valpara\'iso;\\
Casilla 110-V, Valpara\'iso, Chile}
\begin{abstract}
\noindent The so called number of  hadron-nucleus collisions
$n_{coll}(b)$ at impact parameter $b$, and its integral value
$N_{coll}$, which are used to normalize the measured fractional
cross section of a hard process, are calculated  within the
Glauber-Gribov theory including the effects of nucleon short-
range correlations.
%which lead to a modification of the
% hadron-nucleus thickness function
%$T^h_{A}(b)$.
The Gribov inelastic shadowing
corrections are summed to all orders
 by employing the dipole representation.
Numerical calculations are performed  at the
energies of the BNL Relativistic Heavy Ion Collider (RHIC) and CERN Large Hadron Collider (LHC). We found that whereas  the
Gribov corrections generally increase the value of $N_{coll}$,
the inclusion of nucleon correlations, acting in the opposite
directions, decreases it by a comparable amount. The interplay
of the two effects varies with the value of the impact parameter.
\end{abstract}
\vskip 1.5cm
\date{\today}
\maketitle

\section{Introduction}
Much progress has been made recently improving the Glauber
treatment \cite{glauber} of high energy hadron-nucleus and
nucleus-nucleus scattering. From one side, Gribov inelastic
shadowing (IS) corrections \cite{gribov} have been summed up to
all orders
 within the light-cone dipole approach
(see e.g. Refs. \cite{KLZ}-\cite{boris2}). From the other side,
the effects of short range correlations (SRC) on high-energy
scattering has been revisited within realistic approaches
\cite{totalnA}-\cite{nashboris}. In Ref. \cite{nashboris}, the
effects of Gribov IS and SRC on diffractive large rapidity gap
(LRG) processes  and total cross sections have been thoroughly
analyzed. A particular motivation for precise calculations of
these cross sections is  the possibility to use them for
normalization of other channels, which is sometimes a difficult
task. Here such an analysis is extended to the calculation of the
so-called number of collisions $n_{coll}(b)$ and its integral
$N_{coll}$, two quantities that  are  used to normalize the
fractional cross section of hard processes in proton-nucleus and
nucleus-nucleus collisions. Our paper is organized as follows: in
Sec ~\ref{sec:due} the basic elements necessary to treat SRC and
Gribov IS in high-energy hadron-nucleus scattering  are presented;
the definition of the number of collision and their specific form
within our approach, which goes beyond the Glauber model, are
illustrated in Sec ~\ref{sec:tre}; the Results of Calculations are
presented in Sec ~\ref{sec:quattro} and the
summary and conclusions are given in Sec ~\ref{sec:cinque}.\\

\section{Glauber approach, Gribov inelastic shadowing and short range correlations}
\label{sec:due} As is well known, within the  Glauber approach
\cite{glauber}, the elastic hadron-nucleus amplitude reads as
follows
%%%%%%%%%%%%%%%%%%%%%%%%%%%%%%%%%%%%%%%%%%%%%%%%%%%%%%%%%%%%%%%EQUATION
\begin{widetext}
\beq
 \Gamma^{hA}(\Vec b)= \bigg\{1 - \bigg[ 1 -\int\prod_{j=1}^{A}
 d^{3}\,r_j\,\Gamma_j^{hN}(\Vec b -\Vec l_{j})\,
 \delta \Bigl(\sum_{i=1}^A \Vec r_{i}\Bigr)\,
|\psi_0(\Vec r_{1}\dots \Vec r_{A})|^2
\bigg]\bigg\},
\label{total}
\eeq
\end{widetext}
where $\{{\bf r}_i\}=\{{\it{\bf l}}_i, z_i\}$ stands for  a set of
$(A-1)$ linearly independent intrinsic coordinates.
This amplitude depends upon  the modulus squared of the ground-state
 nuclear wave function $|\psi_{0}(\Vec r_1,\dots,\Vec
r_A)|^2 $,
  which, as is now  widely accepted,  exhibits a complex
correlation structure. Thus, the evaluation of various cross
sections accounting for of all possible two-, three-, \dots,
many-body correlations, can be performed either by Monte Carlo
many-fold numerical integration, or by using the following
expansion of  $|\psi_{0}(\Vec r_1,\dots,\Vec r_A)|^2 $  in terms
of density matrices (see Refs.\cite{glauber, foldy})
%%%%%%%%%%%%%%%%%%%%%%%%%%%%%%%%%%%%%%%%%%%%%%%%%%%%%%%%%%%%%%%%%EQUATION
\begin{widetext}
\beq
\left|\,\psi_o(\Vec{r}_1,...,\Vec{r}_A)\,\right|^2=\prod_{j=1}^A\,\rho_1(\Vec{r}_j)
\,+\,\sum_{i<j}\,\Delta(\Vec{r}_i,\Vec{r}_j)\hspace{-0.1cm}\prod_{k\neq (i,j)}\rho_1(\Vec{r}_k)\,+
\hspace{-0.5cm}\sum_{(i<j)\neq(k<l)}\hspace{-0.5cm}\Delta(\Vec{r}_i,\Vec{r}_j)
\,\Delta(\Vec{r}_k,\Vec{r}_l)\hspace{-0.3cm}\prod_{m\neq (i,j,k,l)}\hspace{-0.3cm}
\rho_1(\Vec{r}_m)
\,+\,\dots\,.
\label{psiquadro}
\eeq
\end{widetext}
%%%%%%%%%%%%%%%%%%%%%%%%%%%%%%%%%%%%%%%%%%%%%%%%%%%%%%%%%%%%%%%%%%%%%%%%%%%%%%%%%%%%%%
\vskip 1cm \noindent Here
$\Delta(\Vec{r}_i,\Vec{r}_j)=\rho_2(\Vec{r}_i,\Vec{r}_j)
-\rho_{1}(\Vec{r}_i)\rho_{1}(\Vec{r}_j)$ is the \textit{two-body
contraction} satisfying  the sequential condition $\int
d\Vec{r}_j\,\Delta(\Vec{r}_i,\Vec{r}_j)\,=0$, and
$\rho_{1}(\Vec{r}_i)$ and  $\rho_2(\Vec{r}_i, \Vec{r}_j)$ are the
one- and two-body density matrices, normalized to 1 and obtained
from the general equation \, $\rho_n(\Vec{r}_i,
\Vec{r}_2\dots\Vec{r}_n)= \int \left|\psi_o(\Vec r_{1},\Vec
r_{2},\dots,\Vec{r}_A)\,\right|^2 \prod\displaylimits_{n+1}^A
d\Vec{r}_i $. Note that  in  Eq. (\ref{psiquadro}) the  higher
order terms, not explicitly displayed,  include all possible
products of unlinked two-body contractions, contributing to
two-nucleon correlations, all possible products of unlinked
three-body contractions, describing three-nucleon correlations,
and so forth. Most Glauber-like calculations are based upon the
{\it single-density approximation}, consisting in disregarding all
terms of the expansion (\ref{psiquadro}) except the first one,
i.e.
 $|\psi_{0}(\Vec r_1,\dots,\Vec
r_A)|^2 \approx\prod_{j=1}^{A}\rho(\Vec r_j)$. In the present
paper, we go beyond such an approximation by considering higher
order terms in Eq. (\ref{psiquadro}), and will denote all
quantities calculated within the single-density approximation by a
superscript {\it gl}.
%\vskip-1.2cm
Using the Glauber approach in the treatment of high-energy
hadron-nucleus and nucleus-nucleus scattering is a common
practice;  however it should be stressed that in  this way neither
LRG processes nor intermediate diffractive  hadron dissociation
(Gribov IS) can be explicitly evaluated. It has been shown
\cite{KLZ} that the light-cone dipole is an eigenstate of
high-energy hadron-nucleon interaction, and proper approaches have
been developed  to calculate LRG processes and Gribov IS to all
orders within the single-density approximation (see, e.g., Refs
\cite{boris1}- \cite{boris2}). Recently, the role played by the
correlation terms in Eq. (\ref{psiquadro}) has been revisited,
showing \cite{totalnA} that  two-nucleon correlations (the effects
of higher-order
 correlations being negligible) increase the total neutron-nucleus cross section at high energies,
 making the nucleus more opaque,
  in the opposite direction  of Gribov IS  which increases nuclear transparency.
However in Ref. \cite{totalnA} Gribov IS has been considered
 only at the   lowest order  \cite{karmanov}. A coherent evaluation of IS and SRC
 on the total, $\stota$, elastic,
$\sela$, quasi elastic, $\sqelha$, inelastic, $\sinha$, and
diffractive dissociation hadron-nucleus inclusive cross sections,
has been presented in Ref. \cite{nashboris}, confirming  the
opposite roles played by SRC and IS.
%\vskip -0.35cm
Following the formalism of Ref. \cite{nashboris}, in the
present paper
we further analyze the effects of SRC  and IS by considering
 inelastic hadron-nucleus  collisions at high
 energies.
\section{Number of collisions $n_{coll}(b)$ and $N_{coll}$}
 \label{sec:tre}
 \subsection{Glauber approach and short range correlations}

 We are going to consider  the normalization factor
 that is used to obtain the nucleus to nucleon ratio of  cross sections
 of a hard reaction (high $p_T$, Drell-Yan, heavy flavor, etc), namely
%%%%%%%%%%%%%%%%%%%%%%%%%%%%%%%%%%%%%%%%%%%%%%%%%%%%%%%EQUATION
 \beq
R_{A/N}^{hard} = \frac{\sigma^{hA}_{hard}}
{A\,\sigma^{hN}_{hard}}\ ,
\label{2}
 \eeq
The absolute value of a hard nuclear cross section is difficult to
measure, and  only the fraction of the total number of inelastic
events $N^{hA}_{hard}/N^{hA}_{in}$ is known. One  has therefore to
normalize  the fraction as follows
%%%%%%%%%%%%%%%%%%%%%%%%%%%%%%%%%%%%%%%%%%%%%%%%%%%%%%%EQUATION
 \beq
R_{A/N}^{hard} = \frac{\sina\,N^{hA}_{hard}}
{A\,\sinhn\,N^{hN}_{hard}}= \frac{1}{N_{coll}}\
\frac{N^{hA}_{hard}} {N^{hN}_{hard}}\,, \label{4}
 \eeq
 %%%%%%%%%%%%%%%%%%%%%%%%%%%%%%%%%%%%%%%%%%%%%%%%%%%%%%%%%%%%%%
 where
 %%%%%%%%%%%%%%%%%%%%%%%%%%%%%%%%%%%%%%%%%%%%%%%%%%%%%%EQUATION
 \beq
N_{coll} = A\,\frac{\sinhn}{\sigma^{hA}_{in}}\,, \label{Ncoll}
 \eeq
 %%%%%%%%%%%%%%%%%%%%%%%%%%%%%%%%%%%%%%%%%%%%%%%%%%%%%%%%%%%%%%
and $\sinhn$ is the inelastic hadron-Nucleon inclusive cross section. Correspondingly,
the number of hard collisions at a given impact parameter $b$
should be normalized as
%%%%%%%%%%%%%%%%%%%%%%%%%%%%%%%%%%%%%%%%%%%%%%%%%%%%%%%%%%%EQUATION
 \beq
R_{A/N}^{hard}(b) = \frac{N^{hA}_{hard}(b)}
{n_{coll}(b)\,N^{hN}_{hard}}\ ,
\label{2a}
 \eeq
%%%%%%%%%%%%%%%%%%%%%%%%%%%%%%%%%%%%%%%%%%%%%%%%%%%%%%%%%%%%%%%%%%
where
%%%%%%%%%%%%%%%%%%%%%%%%%%%%%%%%%%%%%%%%%%%%%%%%%%%%%%%%%%%%EQUATION
\beq
 n_{coll}(b)= \frac{\sinhn\,T_A(b)}{P_{in}(b)},
 %{1-e^{[-\sinhn\,T_A^h(b)]}}
\label{ncoll}
 \eeq
 %%%%%%%%%%%%%%%%%%%%%%%%%%%%%%%%%%%%%%%%%%%%%%%%%%%%%%%%%%%%%%%%%%%%%%%%%%%%%%%%%%%%
with  $T_A(b)=\int d\,z \,\rho_A(|{\bf b}|,z)$  being  the usual
 {\it nuclear} thickness function normalized
 to $A$.
 In Eq.(\ref{ncoll}) the numerator represents the number of possibilities of $hN$ interactions,
and  the denominator, $P_{in}(b)$,
  is the probability for an inelastic interaction to occur at impact
  parameter $b$. Note that if the nuclear transparency decreases,
$\sigma^{hA}_{in}$ increases and  $N_{coll}$ decreases, and viceversa. It has been stressed  \cite{boris1}
 that, when Eq. (\ref{Ncoll}) is  used to normalize
hard data on $hA$ and $AA$, care must be taken in the definition
of both the numerator and the denominator. The reasons are  as
follows. If the detector  can detect all diffractive LRG
processes, the numerator, which is taken from
 experimental data on $hN$ scattering, is  $\sigma^{hN}_{in}=
\sigma^{hN}_{tot}-\sigma^{hN}_{el}$  and includes all
diffractive channels;
 the denominator, which has
to be evaluated within some models of hadron-nucleus scattering,
is, accordingly,  $\sigma_{in}^{hA} =\stha - \selha -
\sqelha$ and  includes all diffractive, LRG processes, except the so called
quasi-elastic scattering,
in which bound nucleons remain intact but the nucleus
 gets excited and decays to fragments. If the detector, on the contrary,
 cannot detect LRG processes, these have to be subtracted from the numerator and
 the denominator of Eq. (\ref{ncoll}).
 Our paper aims at analyzing various
relevant models for the evaluation of the denominator. Let us
start with the Glauber model and  the single-density
approximation, supposing that the detector cannot detect LRG
processes. Within such a framework,
 $P_{in}(b)$ is given by
 %%%%%%%%%%%%%%%%%%%%%%%%%%%%%%%%%%%%%%%%%%%%%%%%%%%%%%%%EQUATION
\beq P_{in}^{Gl}(b)= 1-e^{-\sinhn\,T_A^h(b)},
 \label{probability}
\eeq
%%%%%%%%%%%%%%%%%%%%%%%%%%%%%%%%%%%%%%%%%%%%%%%%%%%%%%%%%%%%%%%%%
where $T_A^h(b)$ is the  {\it hadron-nucleus}  thickness function
%%%%%%%%%%%%%%%%%%%%%%%%%%%%%%%%%%%%%%%%%%%%%%%%%%%%%%%%%EQUATION
 \beq
T_A^h({ b})= \frac{2}{\sigma_{tot}^{hN}}\,\int\,d\,^2\,l\, {\text
Re} \Gamma^{hN}(\Vec l) T_A({\Vec b}-
 {\Vec l}),
 \label{thicknessN}
 \eeq
 %%%%%%%%%%%%%%%%%%%%%%%%%%%%%%%%%%%%%%%%%%%%%%%%%%%%%%%%%%%%%%%%
with normalization $\int d^2b\, T_A^h(b)=A$, and
  the inclusive inelastic hadron-nucleus cross section is
  %%%%%%%%%%%%%%%%%%%%%%%%%%%%%%%%%%%%%%%%%%%%%%%%%%%%%%%EQUATION
\beq
\sigma_{in}^{hA} =\stha - \selha -
\sqelha=\int d^2b\, \left[1-e^{-\sinhn\,T_A^h(b)}\right].
\label{sigmaprod}
\eeq
%%%%%%%%%%%%%%%%%%%%%%%%%%%%%%%%%%%%%%%%%%%%%%%%%%%%%%%%%%%%%%%%%
The total number of
collisions will then read
%%%%%%%%%%%%%%%%%%%%%%%%%%%%%%%%%%%%%%%%%%%%%%%%%%%%%%%%%EQUATION
\beq
N_{coll}^{Gl}=
%\frac{\int d^2b N_{p}(b)}{\int d^2b P_{NA}(b)}= A
 A \frac{\sinhn}{\int d\,^2\,b\,\left[ 1-e^{-\sinhn\,T_A^h(b)}\right]},
\label{Ncollgl}
\eeq
%%%%%%%%%%%%%%%%%%%%%%%%%%%%%%%%%%%%%%%%%%%%%%%%%%%%%%%%%%%%%%%%%%
where  the denominator includes all diffractive LRG processes, but
misses  the effects from SRC and Gribov IS. SRC can readily be
implemented  in Eq. (\ref{Ncollgl}) but the inclusion of Gribov IS
is no easy theoretical task since the Glauber model, which
 is a single-channel approximation,
cannot be applied, and more involved multichannel approach or the
dipole representation should be used
\cite{boris1,boris2,nashboris}. In this paper we have calculated
$n_{coll}$, Eq. (\ref{ncoll}), and its integral $N_{coll}$, Eq.
(\ref{Ncoll}),  with four different approaches for the evaluation
 of $\sigma_{in}^{hA}$: (i) the plain Glauber model within the single-density
 approximation given by Eq. (\ref{sigmaprod}); (ii)
 the improved  Glauber model which includes the effects of SRC; and  (iii) the model in which,  besides SRC,
 also  LRG processes and Gribov IS, calculated within the dipole approach, are included in $\stha$, $\selha$, and
$\sqelha$; iv) finally, since frequently experiments select only
 events with particle production
at central rapidities,  missing LRG diffractive channels, we have
modified accordingly \cite{boris1} the model iii) by subtracting
 the cross sections of single and double diffraction  from  $\sigma^{hN}_{in}$ in the numerators and
the contribution of LRG processes from the denominator.

Let us briefly present the theoretical background underlying the
above program  (see Ref. \cite{nashboris}for more details).
Concerning  the effects of SRC
on the various cross sections,
 it has been shown \cite{totalnA}
 that their inclusion is equivalent
 to the following  modification of the hadron-nucleus  thickness
 function,
%%%%%%%%%%%%%%%%%%%%%%%%%%%%%%%%%%%%%%%%%%%%%%%%%%%%%%%%%EQUATION
\beq
T_A^h(b)\Rightarrow \widetilde T_A^h(b)= T_A^h(b) -\Delta T_A^h(b),
\label{tildeti}
\eeq
%\end{widetext}
%%%%%%%%%%%%%%%%%%%%%%%%%%%%%%%%%%%%%%%%%%%%%%%%%%%%%%%%%%%%%%%%%%
with the correlation correction $\Delta T_A^h(b)$ given by
%%%%%%%%%%%%%%%%%%%%%%%%%%%%%%%%%%%%%%%%%%%%%%%%%%%%%%%%%%EQUATION
\begin{widetext}
 \beq
  \Delta T_A^h(b)=
  \frac{1}{\sthn}
  \int d^2l_1\,d^2l_2\,
  \Delta^\perp_A(\Vec l_1,\Vec l_2)
  \Re \Gamma^{pN}(\Vec b-\Vec l_1)\,
  \Re \Gamma^{pN}(\Vec b-\Vec l_2),
  \label{55}
  \eeq
  \end{widetext}
%%%%%%%%%%%%%%%%%%%%%%%%%%%%%%%%%%%%%%%%%%%%%%%%%%%%%%%%%%EQUATION
  where
  \beq
  \Delta^\perp_A(\Vec l_1,\Vec l_2)=
  A^2 \int\limits_{-\infty}^\infty dz_1
  \int\limits_{-\infty}^\infty dz_2\,
   \Delta(\Vec r_1,\Vec r_2)
   \label{55a}
   \eeq
%%%%%%%%%%%%%%%%%%%%%%%%%%%%%%%%%%%%%%%%%%%%%%%%%%%%%%%%%%%%%%%%%%%
is the  transverse two-nucleon contraction. In Eq. (\ref{55}) two
nucleon correlations, represented by all possible products of
two-body contractions  linked to all possible products of two
$\Gamma^{pN}(\Vec b-\Vec l_i)$, are exactly summed up. It should
be stressed that the inclusion of SRC does not affect the
numerator of Eq. (\ref{Ncollgl}), but it will change  the
probability $P_{in}(b)$, which is still given by Eq.
(\ref{probability}) but with $T_A^h(b)$ Eq. (\ref{thicknessN})
replaced by $\widetilde T_A^h(b)$ Eq. (\ref{tildeti}).
\subsection{Adding Gribov corrections via  light-cone dipoles}
The effects of both Gribov IS and SRC have been considered as in
Refs. \cite{boris1,boris2,nashboris} within the light-cone dipole
approach.
 The dipole
representation for the amplitude of hadronic interactions allows
one to sum up the Gribov inelastic corrections to all orders. If
the collision energy is  high enough to keep the dipole size
''frozen'' by Lorentz time delation during propagation through the
nucleus, the calculations are much simplified. The key ingredients
of the approach are the universal dipole-nucleon cross section
$\sigma_{dip}(r_T,s)$ ($r_T$ is the transverse dimension of the
${\bar q} q$ dipole and $s$ is the energy)
 and the light-cone wave function of the projectile
hadron  $\Psi_N({\Vec r}_1,{\Vec r}_2,{\Vec r}_3)$ \cite{boris2}. In this paper, as in Ref. \cite{nashboris},
 the models for both quantities
 have been taken from Ref. \cite{boris2}. Both models employ
the saturated shape of the dipole cross section and differ only by
modeling the proton wave function. For the sake of illustration,
let us consider  $\sigma_{tot}^{pA}$. Taking into account both
Gribov IS and SRC,  one gets \cite{nashboris}
%%%%%%%%%%%%%%%%%%%%%%%%%%%%%%%%%%%%%%%%%%%%%%%%%%%%%%%%%%%%%%%%%%%%%%%%%%%%%EQUATION
\beqy
\stap =
 2\int d^{\,2}b
\left[1-\left\la e^{-\frac{1}{2}\sigma_{dip}(r_T,s)
{\widetilde T}^{dip}_A(b,r_T,\alpha)}\right\ra\right]
\label{total1}
 \eeqy
%%%%%%%%%%%%%%%%%%%%%%%%%%%%%%%%%%%%%%%%%%%%%%%%%%%%%%%%%%%%%%%%%%%%%%%%%%%%%%%%%%%%%
\noindent where the average is over the transverse size of the dipole $q\bar q$ and the fractional
 light-cone momentum $\alpha$, i.e., for a generic function $f(r_T,
 \alpha)$,
%%%%%%%%%%%%%%%%%%%%%%%%%%%%%%%%%%%%%%%%%%%%%%%%%%%%%%%%%%%%%%%%%%%%%%%%%%%%%EQUATION
 \beq
\la  f(r_T,\alpha)\ra \equiv \int\limits_0^1
d\alpha\int\limits_0^{\infty} d^{\,2}r_T\,
|\Psi_N(r_T,\alpha)|^2\,f(r_T,\alpha).
 \eeq
% \end{widetext}
 %%%%%%%%%%%%%%%%%%%%%%%%%%%%%%%%%%%%%%%%%%%%%%%%%%%%%%%%%%%%%%%%%%%%%%%%%%%%%%%%%%%%%
The correlated dipole tickness function reads
%, due to SRC, one has
 %%%%%%%%%%%%%%%%%%%%%%%%%%%%%%%%%%%%%%%%%%%%%%%%%%%%%%%%%%%%%%%%%%%%%%%%%%%%%EQUATION
% \begin{widetext}
 \beq
  {\widetilde T}^{dip}_A(b,r_T,\alpha)= T^{dip}_A(b,r_T,\alpha) + \Delta T^{dip}_A(b,r_T,\alpha)
  \eeq
%  \end{widetext}
%%%%%%%%%%%%%%%%%%%%%%%%%%%%%%%%%%%%%%%%%%%%%%%%%%%%%%%%%%%%%%%%%%%%%%%%%%%%%%%%%%%%%%
  with
%%%%%%%%%%%%%%%%%%%%%%%%%%%%%%%%%%%%%%%%%%%%%%%%%%%%%%%%%%%%%%%%%%%%%%%%%%%%%%EQUATION
%\begin{widetext}
  \begin{gather}
T^{dip}_A(b,r_T,\alpha)= \nonumber \\
=\frac{2}{\sigma_{dip}(r_T)}\int d^{\,2}l \Re \Gamma^{dip}(\Vec
l,\Vec r_T,\alpha)T_A(\Vec b-\Vec l) \label{214}
\end{gather}
%\end{widetext}
%%%%%%%%%%%%%%%%%%%%%%%%%%%%%%%%%%%%%%%%%%%%%%%%%%%%%%%%%%%%%%%%%%%%%%%%%%%%%%%%%%%%%%%
and
%%%%%%%%%%%%%%%%%%%%%%%%%%%%%%%%%%%%%%%%%%%%%%%%%%%%%%%%%%%%%%%%%%%%%%%%%%%%%%%EQUATION
\begin{widetext}
 \beqy
  \Delta T_A^{dip}(b,r_T,\alpha)=
  \frac{1}{\sigma_{dip}(r_T)}
  \int d^{\,2}l_1\,d^{\,2}l_2
  \Delta^\perp_A(\Vec l_1,\Vec l_2)
  \Re \Gamma^{dip}(\Vec b-\Vec l_1,r_T,\alpha)
  \Re \Gamma^{dip}(\Vec b-\Vec l_2,r_T,\alpha).
  \label{500}
  \eeqy
  \end{widetext}
%%%%%%%%%%%%%%%%%%%%%%%%%%%%%%%%%%%%%%%%%%%%%%%%%%%%%%%%%%%%%%%%%%%%%%%%%%%%%%%%%%%%%%%%
Here $\Delta^\perp_A(\Vec l_1,\Vec l_2)$ is given by Eq.
(\ref{55a}) and the partial dipole-nucleon amplitude $\Re
\Gamma^{dip}$ is given by Eq. (41) of Ref. \cite{nashboris}. Using
the above equations, the total cross section  takes the form,
%%%%%%%%%%%%%%%%%%%%%%%%%%%%%%%%%%%%%%%%%%%%%%%%%%%%%%%%%%%%%%%%%%%%%%%%%%%%%%%%EQUATION
%    \begin{widetext}
  \beq
\stap = 2\int d^2b\,
\Biggl\{1-e^{{1\over2}I_A(b)}
\left\la e^{-\frac{1}{2}\sigma_{dip}
T^h_A(b)}\right\ra\Biggr\}
\label{530}
 \eeq
%\end{widetext}
%%%%%%%%%%%%%%%%%%%%%%%%%%%%%%%%%%%%%%%%%%%%%%%%%%%%%%%%%%%%%%%%%%%%%%%%%%%%%%%%%%%%%%%%
where the quantity $I_A(b)$ contains the effects from SRC
\cite{nashboris}.
\section{Results of calculations}
\label{sec:quattro} We have calculated  $n_{coll}$ and $N_{coll}$
in proton-nucleus scattering at BNL Relativistic Heavy Ion
Collider (RHIC) and CERN Large Hadron Collider (LHC)LHC energies
within the following approaches: i) the Glauber model with
single-density approximation
%%%%%%%%%%%%%%%%%%%%%%%%%%%%%%%%%%%%%%%%%%%%%%%%%%%%%%%%%%%%%%%%%%%%%%%%%%%%%%EQUATION
\beqy
 n_{coll}^{Gl}(b)= \frac{\sinhn\,T_A(b)}{1-e^{-\sinhn\,T_A^h(b)}};
 \label{Pglauber}
 \eeqy
 ii) the Glauber model plus SRC
 %%%%%%%%%%%%%%%%%%%%%%%%%%%%%%%%%%%%%%%%%%%%%%%%%%%%%%%%%%%%%%%%%%%%%%%%%%%%%%EQUATION
 \beqy
n_{coll}^{Gl+SRC}(b)=
\frac{\sinhn\,{T}_A(b)}{1-e^{-\sinhn\,{\widetilde T}_A^h(b)}};
 \label{PglauberSRC}
 \eeqy
 iii) the Glauber model plus SRC and  Gribov IS corrections,
  using the results of Ref. \cite{nashboris}.
 In this case, we have also considered  the possibility
 that the detector misses  LRG processes  which, therefore,
 have
 to be subtracted from the numerator and the denominator of
 Eq. (\ref{PglauberSRC}),
 arriving at
%%%%%%%%%%%%%%%%%%%%%%%%%%%%%%%%%%%%%%%%%%%%%%%%%%%%%%%%%%%%%%%%%%%%%%%%%%%%%%EQUATION
\beqy
 n_{coll}^{Gl+SRC+IS}(b)=\frac{(\sinhn-\sigma_{diff}^{hN})\,{T}
_A(b)}{P_{in}(b)},
\label{PglauberSRCIS}
 \eeqy
%%%%%%%%%%%%%%%%%%%%%%%%%%%%%%%%%%%%%%%%%%%%%%%%%%%%%%%%%%%%%%%%%%%%%%%%%%%%%%%EQUATION
where the probability $P_{in}(b)$ is given by
\beq
P_{in}(b)=
\frac{d\sigma_{tot}^{hA}}{d^2b}-\frac{d\sigma_{el}^{hA}}{d^2b}-
\frac{d\sigma_{diff}^{hA}}{d^2b}-\frac{d\sigma_{qel}^{hA}}{d^2b}-\frac{d\sigma_{qsd}^{hA}}{d^2b},
\label{p-in}
\eeq
 with
 %%%%%%%%%%%%%%%%%%%%%%%%%%%%%%%%%%%%%%%%%%%%%%%%%%%%%%%%%%%%%%%%%%%%%%%%%%%EQUATION
  \beqn
{1\over2}\,\frac{d\sigma_{tot}^{hA}}{d^2b}= 1-e^{{1\over2}I_A(b)}
\left\la e^{-\frac{1}{2}\sigma_{dip} T^h_A(b)}\right\ra,
\label{p-in-a}
\eeqn
\beqn
\frac{d(\sigma_{el}^{hA}+\sigma_{diff}^{hA})}{d^2b} =
\frac{d\sigma_{tot}^{hA}}{d^2b} - 1+e^{I_A(b)} \left\la
e^{-\sigma_{dip}T^h_A(b)}\right\ra,
\label{p-in-b}\eeqn
\begin{gather}
\frac{d(\sigma_{qel}^{hA} + \sigma_{qsd}^{hA})}{d^2b}=\nonumber \\
\Biggl\la e^{-
\sigma_{dip}T^h_A(b)}
\left\{e^{\tilde I_A(b)} e^{\frac{\sigma_{dip}^2T^h_A(b)} {16\pi
B_{el}}} -e^{I_A(b)}\right\} \Biggr\ra.
 \label{p-in-c}
 \end{gather}
%%%%%%%%%%%%%%%%%%%%%%%%%%%%%%%%%%%%%%%%%%%%%%%%%%%%%%%%%%%%%%%%%%%%%%%%%%%%%%%%%%%%%%
Here the quantities $I_A(b)$ and $\tilde I_A(b)$, providing the effects from SRC are
\begin{widetext}
\beqn
I_A(b)=\Bigl\la \sigma_{\bar qq}(r_T)\, \Delta T^{\bar
qq}_A(b,r_T,\alpha)\Bigr\ra
=\left[\sigma^{pN}_{el}+\sigma^{pN}_{sd}\right] \times \int
d^2\delta\,\exp\left[-\frac{\delta^2}{4B(s)+R_0^2(s)/2}\right]\,
\Delta^\perp_A(\vec \delta,b),
\label{520}
 \eeqn
%\begin{widetext}
 \begin{table}[!t]
\begin{center}
$p-^{208}Pb$
\end{center}
\begin{center}
{GLAUBER}
\end{center}
\begin{tabular*}{1.0\textwidth}{@{\extracolsep{\fill}}c| c c c c c c c}\hline\hline
%& & & GLAUBER & & & \\ \hline
& $\sigma_{in}^{pN}\:[mb]$ & $\sigma_{tot}^{pA}\:[mb]$ &
$\sigma_{el}^{pA} \:[mb]$ & $\sigma_{qel}^{pA}
\:[mb]$ & $\sigma_{in}^{pA}\:[mb]$& $N_{coll}$\\
\hline RHIC &42.1  &3297.6 &1368.4 &66.0 &1863.2&4.70 \\
\hline LHC  &68.3  & 3850.6 &1664.8 &121.0 &2064.8&6.88\\
\hline \hline
 \end{tabular*}
%    \end{table}
%%%%%%
%%%%% GLAUBER + SRC
\begin{center}
{GLAUBER+SRC}
\end{center}
%\vskip-0.5cm
%\begin{table}[!h]
 \begin{tabular*}{1.0\textwidth}{@{\extracolsep{\fill}}c| c c c c c c c}\hline\hline
%& & & GLAUBER+SRC & & & \\ \hline
& $\sigma_{in}^{pN}\:[mb]$ & $\sigma_{tot}^{pA}\:[mb]$ &
$\sigma_{el}^{pA} \:[mb]$ &
$\sigma_{qel}^{pA}\:[mb]$ & $\sigma_{in}^{pA}\:[mb]$  & $N_{coll}$\\
\hline RHIC &42.1  &3337.6 &1398.1 &58.5 &1881.0 &4.65\\
\hline LHC  &68.3  & 3885.8& 1690.5&112.6 & 2082.7&6.82\\
\hline \hline
 \end{tabular*}
    %\end{table}
%%%%% GLAUBER + SRC + GRIBOV
\begin{center}
{GLAUBER+SRC+GRIBOV IS }
\end{center}
%\vskip-0.5cm
%\begin{table}[!h]
 \begin{tabular*}{1.0\textwidth}{@{\extracolsep{\fill}}c| c c c c c c c}\hline\hline
%& & & GLAUBER+SRC+GRIBOV($q-2q$)& & & \\ \hline
& $\sigma_{in}^{pN}$ ($\sigma_{in}^{pN}-\sigma_{diff}^{pN})\:[mb]$ &
$\sigma_{tot}^{pA}\:[mb]$ & $\sigma_{el}^{pA}$ ($\sigma_{el}^{pA}+\sigma_{diff}^{pA})
\:[mb]$ & $\sigma_{qel}^{pA}$ ($\sigma_{qel}^{pA}+\sigma_{qsd}^{pA})\:[mb]$ & $\sigma_{in}^{pA}\:[mb]$  & $N_{coll}$\\
\hline RHIC &42.1 (30.0) &3228.1 &1314.0 (1331.0) &72.0 (74.4) & 1842.1 (1823.0) & 4.75(3.42)\\
\hline LHC  &68.3 (56.3)  &3833.3 &1655.7 (1658.0) &113.4 (111.3) & 2064.2 (2064.0) &6.88(5.67) \\
\hline \hline
 \end{tabular*}
\caption{$N_{coll}$ Eq. (\ref{Ncoll}) in $p-^{208}Pb$ scattering
at RHIC and LHC energies calculated by integrating the numerator
and denominator of Eqs. (\ref{Pglauber}), (\ref{PglauberSRC}), and
(\ref{PglauberSRCIS}). In the latter case (GLAUBER+SRC+GRIBOV)the
values  in parentheses correspond to to the full Eqs.
(\ref{PglauberSRCIS})-(\ref{p-in-c}) and represent the case when
the detector misses LRG events, whereas the other values have been
obtained by omitting the diffractive cross section
$\sigma_{diff}^{hN}$ from
 Eq. (\ref{PglauberSRCIS}) and  $d\sigma_{diff}^{hA}/d^2b$
and $d\sigma_{qsd}^{hA}/d^2b$ from Eq.(\ref{p-in}),
 to describe processes when LRG events are detected.
}.
\end{table}

%%%%%%%%%%%%%%%%%%%%%%%%%%%%%%%%%%%%%%%%%%%%%%%%%%%%%%%%%%%%%%%%%%%%%%%%%%%%%%%%%
and
%%%%%%%%%%%%%%%%%%%%%%%%%%%%%%%%%%%%%%%%%%%%%%%%%%%%%%%%%%%%%%%%%%%%%%%%%EQUATION
\beq
\tilde I_A(b) \approx
\left[\frac{\sigma_{tot}^{pN}-\sigma_{el}^{pN}-\sigma_{sd}^{pN}}{\sigma_{tot}^{pN}}\right]^2
I_A(b).
 \label{620}
\eeq
\end{widetext}
In these equations, $\sigma_{diff}^{hN}$ includes the diffraction
dissociation of the projectile and   target hadrons,
$\sigma_{diff}^{hA}$ includes the diffraction dissociation of the
projectile ($\sigma_{sd}$ of Ref. \cite{nashboris}), and
$\sigma_{qsd}^{hA}$ the diffraction dissociation of both the
projectile hadron and the target nucleus (see Ref.
\cite{nashboris} for more details and notations). Equations
(\ref{PglauberSRCIS})-(\ref{p-in-c}) have been calculated
considering and omitting  the diffractive cross sections
$\sigma_{diff}^{hN}$, $\sigma_{diff}^{hA}$ and
$\sigma_{qsd}^{hA}$; the former case  corresponds to a probability
$P_{in}(b)$, which, when integrated, yields the non diffractive
cross section
%$\sigma_{in}^{hA} =\stha - \selha -
%\sqelha$
 and represents the case when the detector
is insensitive
 to LRG channels (e.g. in the experiments STAR \cite{star},
PHENIX \cite{phenix} and PHOBOS \cite{phobos} at RHIC);  the
second case corresponds to the assumption that the detector can
detect the  LRG processes. In the calculations we used the
experimental cross sections $\sinn=42 \,mb$ and  $\sinn=68.43
\,mb$ for RHIC and LHC energies, respectively.  As for the nuclear
quantities,  realistic one- and two- body densities and
correlation
 functions from Ref. \cite{ACMprl}  have been adopted (see Ref. \cite{nashboris} for details).
 Notice that, eventually,  by definition, the nuclear thickness
 function $T_A(b)$ has to be used in the numerators of Eqs.
 (\ref{Pglauber})-(\ref{p-in}) since, as already pointed out, the number of opportunities to perform a hard collision
depends only upon the
 single-particle density; the denominator, on the contrary,  is affected  by SRC, since these  couple two-body contractions
 with products of two profile $\Gamma$, according
 to Eqs. (\ref{tildeti}) and (\ref{500}).
 %\begin{widetext
%%%%%%%%%%%%%%%%%%%%%%%%%%%%%%%%%%%%%%%%%%%%%%%%%%%%TABLE
%\end{widetext}
The results of calculations pertaining to $p-^{208}Pb$ collisions
are presented in Table I. It can be seen that SRC increase the
value of $\sigma_{in}^{NA}$ and, correspondingly, decrease the
values of $N_{coll}$; at the same time, Gribov IS increases them
back to values near  to the Glauber ones. Similar results are
obtained for other nuclei. The effects of both SRC and Gribov IS
amounts to a few percent, in agreement with the results of the
 calculation of deuteron-gold scattering \cite{boris1}.
Our results, together with the results previously obtained for LRG
cross sections, generally show that, although the separate effects
of Gribov IS and SRC on the  quantities we have considered can be
appreciable, they tend to act in the opposite directions, making
the final result
 similar to the one obtained within the Glauber approximation. However, it should also be pointed out that when the detector misses the LRG processes
 (the values in parentheses),  $N_{coll}$ turns out to be appreciably lower than the Glauber value.
 \section{Summary and conclusions}\label{sec:cinque}
In this paper, we continued our analysis of the effects of
short-range correlations and Gribov inelastic
  shadowing on high-energy scattering processes
  by considering the number of collisions, the quantity that is used to
 normalize the fractional cross
section of hard processes in proton-nucleus and nucleus-nucleus
collisions. The short-range correlations are treated within a
realistic many-body framework, and the effects of Gribov inelastic
shadowing are summed up to all orders by  the light-cone dipole
approach. The numerical results confirm the opposite role played
by correlations  and Gribov inelastic shadowing and, once again,
point at the necessity of inclusion of both effects whenever a
precise analysis of experimental data is required.
%%%%%%%%%%%%%%%%%%%%%%%%%%%%%%%%%%%%%%%%%%%%%%%%%%%%%%%%%%%%%%%%%%%%%%%% END TABLE

\end{document}